\documentclass[%
 reprint,
nofootinbib,
 amsmath,amssymb,superscriptaddress,
 aps,
]{revtex4-1}

\usepackage{graphicx}
\usepackage{dcolumn}
\usepackage{bm}

\usepackage[utf8]{inputenc}
\usepackage{amssymb}
\usepackage{amsmath}
\usepackage[colorlinks=true,linkcolor=black, citecolor=blue, urlcolor=blue]{hyperref}
\usepackage{appendix}
\usepackage{booktabs}
\usepackage{tabularx}
\usepackage{ltablex}
\usepackage{xcolor}
\usepackage{longtable}
\usepackage{lipsum}
\usepackage{soul}
\usepackage{braket}
\usepackage{footnote}
\usepackage{xr}
\usepackage{soul}

\usepackage{mathtools}

\begin{document}

\preprint{APS/123-QED}

\title{Quantifying and suppressing ranking bias in a large citation network}
%
%

\author{Giacomo Vaccario}
\affiliation{Chair of Systems Design, ETH Zurich, 
8092 Zurich, Switzerland}
\author{Mat{\'u}{\v{s}} Medo}
\affiliation{Department of Physics, University of Fribourg, 1700 Fribourg, Switzerland}
\affiliation{Department of Radiation Oncology, Inselspital, Bern University Hospital and University of Bern, 3010 Bern, Switzerland}
\author{Nicolas Wider}
\affiliation{Chair of Systems Design, ETH Zurich, 
8092 Zurich, Switzerland}
\author{Manuel Sebastian Mariani}
\email{manuel.mariani@unifr.ch}
\affiliation{Department of Physics, University of Fribourg, 1700 Fribourg, Switzerland}


\begin{abstract}
It is widely recognized that citation counts for papers from different fields cannot be directly compared because different scientific fields adopt different citation practices.
Citation counts are also strongly biased by paper age since older papers had more time to attract citations. 
Various procedures aim at suppressing these biases and give rise to new normalized indicators, such as the relative citation count. 
We use a large citation dataset from Microsoft Academic Graph and a new statistical framework based on the
Mahalanobis distance to show that the rankings by well known indicators, including the relative citation count and Google's PageRank score, are significantly biased by paper field and age. 
We propose a general normalization procedure motivated by the $z$-score which produces much less biased rankings when applied to citation count and PageRank score.
\end{abstract}

\maketitle

\section{Introduction}

Paper citation count itself and various quantities derived from it are used as influential indicators of research impact \citep{hirsch2005index,garfield2006history}. 
At the same time, it is well known that the cumulative number of citations received by academic publications strongly depend on paper age and field~\citep{schubert1986relative, vinkler1986evaluation}.
Old papers have had more time to acquire citations than recent ones, and their  advantage is further enhanced by the preferential attachment mechanism \citep{price1976general, newman2009first,medo2011temporal}.
On the other hand, different academic fields adopt very different citation practices (see \cite{bornmann2008citation} for a review on the topic), which results in a strong dependence of the mean number of citations on academic field, as shown in several works (\cite{lundberg2007lifting,radicchi2008universality,bornmann2009universality}, among others).

A natural question arises: how can we "fairly" use citation-based indicators to compare papers from different fields and of different age? 
The problem of comparing papers from different fields is usually referred to as the \emph{field-normalization} problem. Several approaches to address this question have been proposed in the literature (see \cite{waltman2016review} for a recent review). A particularly simple approach is to divide each paper's citation count by the mean number of citations for papers of the same field published in the same year.
The results by \cite{radicchi2008universality} 
suggested that this indicator, called \emph{relative citation count}, produces a ranking that is statistically consistent with the hypothesis of a ranking that is not biased by field and age. 
This finding has been challenged by subsequent works by \cite{albarran2011skewness} and \cite{waltman2012universality}, which leaves the debate on age- and field- normalization procedures still open.

In this article, we analyze a large dataset from Microsoft Academic Graph \citep{sinha2015overview} to show that existing indicators of impact, including the relative citation count, fail to produce rankings that are not biased by age and field. 
To simultaneously assess these biases, we present a new procedure based on the Mahalanobis distance~\citep{Mahalanobis1936distance}. This permits to compare the ranking by a given indicator with those obtained with a simulated unbiased sampling, and hence to quantify the overall ranking bias. 
An analytic result derived in this paper allows us to assess the contribution of each field to the overall ranking bias.
It is worth noticing that while we focus on the biases by age and field, our bias assessment procedure can be easily extended to detect any other kind of information bias.

We also present the first systematic study of the possible bias by field of PageRank score \citep{brin1998anatomy} and of its age-rescaled version introduced by \cite{mariani2016identification}.
The motivation to analyze these network-based indicators comes from the finding that they outperform other metrics in identifying expert-selected milestone papers \citep{mariani2016identification}.
However, the application of PageRank and its variants to academic citation networks focused on datasets composed of papers from a single field~\citep{chen2007finding,walker2007ranking,yao2014ranking,mariani2015ranking,mariani2016identification,zhou2016ranking}.
To our best knowledge, the metric's possible bias by academic field is still unexplored and we are the first authors to address it.

We introduce two novel indicators of impact motivated by the $z$-score: age- and field-rescaled citation count $R^{AF}(c)$ and age- and field-rescaled PageRank $R^{AF}(p)$.
We find that the novel indicators produce paper rankings that are much less biased by age and field than the rankings produced by the other analyzed indicators. 
Nevertheless, also the Mahalanobis distance observed for the new indicators is not statistically consistent with ones obtained for a simulated unbiased process. 
This indicates that the problem of achieving an ideal unbiased ranking of the publications remains open.

The rest of our article is organized as follows: Section 2 describes the analyzed dataset of publications obtained from the Microsoft Academic Graph. 
Section 3 presents existing paper-level impact indicators and reports their bias by scientific field. 
In Section 4, we introduce a rescaling procedure for citation count and PageRank scores motivated by the $z$-score.
In Section 5, we introduce a general procedure to test for any kind of ranking bias, and present its application to assess the field and age bias of the rankings by the indicators studied here.
In Section 6, we conclude by discussing possible limitations of our analysis and future research directions.

\section{Data}  
\label{sec:data}

We analyze a bibliographic dataset which was provided for the KDD Cup 2016\footnote{\url{https://kddcup2016.azurewebsites.net/Data}}. 
This data is a dump of the \emph{Microsoft Academic Graph} (MAG)
and contains more than 126 millions of publications and more than 467 millions citations \citep{sinha2015overview}.
Each publication is also endowed with various properties such as unique ID, publication date, title, journal ID, etc.
We pre-processed the data (details are provided in Appendix \ref{app:kdd}) to remove from the analysis papers with incomplete information, ending up with $N=18\,193\,082$ unique publications and $E=109\,719\,182$ citations.

The MAG has a field classification at paper level \citep{sinha2015overview}. 
In the KDD cup dump, there are $19$ main fields and numerous subfields up to $3$ hierarchical levels of subsubfields. 
However, all the different subfields can belong to several main fields, meaning that each publication can belong to more than one main field.
We use here the field classification at the highest hierarchical level, i.e., we only consider the $19$ main fields.
When calculating the citation count and PageRank score of papers (see Sec.~\ref{sec:short-ex-metric}), we consider the publications that belong to more than one field only once. 
In this way, we do not modify the number of citations that each paper receives and gives, and we do not change the topology of the network on which the PageRank scores are calculated. 
On the other hand, in agreement with \citep{waltman2012universality}, to compute the fields' size (see Tab.~\ref{tab:fields} in Appendix \ref{app:kdd}) and the field-rescaled metrics (see Sec.~\ref{sec:short-ex-metric} and Sec.~\ref{sec:def-new-metrics}), each publication can be considered multiple times in the analysis, once for each field the publication belongs to.
With this, each field is represented by all its publications even if some of these are shared with other fields.

\section{Shortcomings of existing metrics}
\label{sec:short-ex-metric}

We now define the four existing metrics analyzed in this work: citation count $c$, relative citation count $c^{f}$, PageRank score $p$, age-rescaled PageRank score $R^{A}(p)$ (Sec.~\ref{sec:definition}). Furthermore, we show that these metrics are severely biased by scientific field (Sec.~\ref{sec:field_bias}).

\subsection{Definition of existing metrics}
\label{sec:definition}

\paragraph{Citation count, $c$}
The citation count $c_i$ of node $i$ is simply the number of citations received by  paper $i$.
In terms of the citation network's adjacency matrix $\mathbf{A}$ (in a directed network, $A_{ij}=1$ if node $j$ points to node $i$, 
 $A_{ij}=0$ otherwise), we can express the citation count as $c_{i}=\sum_j A_{ij}$.

\paragraph{Relative citation count, $c^f$}
To overcome citation count's bias by paper age and academic field, \cite{radicchi2008universality} defined the \emph{relative citation count} $c^{f}_i$ of paper $i$ as $c^{f}_i:=c_i/\mu^{Y}_{i}(c)$, where $\mu^{Y}_i(c)$ denotes the mean citation count for papers published in the same field and year as paper $i$.
Throughout this paper, we always refer to the $19$ main fields provided in the MAG dataset. 

\paragraph{Page Rank, $p$}
Originally devised by \cite{brin1998anatomy} to rank webpages in the World Wide Web, PageRank algorithm has attracted considerable interest of the scientometrics community.
The rationale behind the application of the metric to citation networks is that not all citations should be counted the same: citations coming from influential papers should count more than citations from obscure articles.
The PageRank scores of papers are usually written in a vector $\mathbf{p}$ defined by the following equation
\begin{equation}
 \mathbf{p}=\alpha \,\mathbf{P}\,\mathbf{p}+(1-\alpha)\mathbf{v},
\label{pr_eq}
 \end{equation}
where $\alpha$ is a constant, $\mathbf{P}$ is the random-walk transition matrix with elements $P_{ij}=A_{ij}/k^{out}_j$, $k^{out}_j=\sum_{l}A_{lj}$ is the number of references in paper $j$, and $\mathbf{v}$ is a uniform teleportation vector with elements $v_{i}=1/N$ for all papers $i$.
Eq.~\eqref{pr_eq} can be interpreted as the stationary equation of a stochastic process on the citation network. 
In this process, a random walker is placed on each paper and he/she either follows a citation edge with probability $\alpha$, or jumps to a randomly chosen paper with probability $1-\alpha$.
When the number of walkers on each paper reaches a stationary value, we obtain the PageRank score of a paper $i$ by calculating the fraction of walker on this paper.
In our analysis, we set $\alpha=0.5$ which is the usual choice in citation networks \citep{walker2007ranking}.


PageRank is based on a static, time-aggregated perspective of the considered network. 
In general, such perspective has been shown to be limiting for the analysis of evolving networks~\citep{scholtes2014causality, mariani2015ranking}. 
While the resulting metric's bias towards old papers has already been studied in the literature \citep{chen2007finding,maslov2008promise,mariani2015ranking,mariani2016identification}, its possible bias by academic field is still unexplored and we address it in Section~\ref{sec:field_bias}.

\paragraph{Age-rescaled Page Rank, $R^{A}(p)$}
To suppress the age bias of PageRank, \cite{mariani2016identification} proposed to rescale the PageRank score by comparing each paper's score with the scores of papers of similar age.
Assuming that the papers are ordered by older to younger, one computes the mean value $\mu^{A}_i(p)$ and the standard deviation $\sigma^{A}_i(p)$ of PageRank scores over $\Delta_p$ papers around $i$, i.e. $j\in[i-\Delta_p/2,i+\Delta_p/2]$.
Consequently, the rescaled PageRank score $R^{A}_i(p)$ of paper $i$ is defined as
\begin{equation}
 R^{A}_i(p)=\frac{p_i-\mu^{A}_{i}(p)}{\sigma^{A}_{i}(p)}.
\end{equation}
\cite{mariani2016identification} applied rescaled PageRank to the network of physics papers published by the American Physical Society journals to show that the resulting ranking is not biased by paper age and, as a result, it allows us to identify seminal publication much earlier than rankings by metrics that are biased against recent papers. 
In the following, we set $\Delta_p=1000$ as in \citep{mariani2016identification}. 

\subsection{Field bias of the existing metrics}
\label{sec:field_bias}
After having described a set of existing metrics, we now apply them to the MAG dataset to show that the rankings that they produce are biased by scientific field.
For a ranking that is not biased by scientific field, the number of top-ranked publications from each field should be proportional to the total number of publications from that field.
In other words, for an unbiased ranking, we expect
\begin{equation}
\mu_f=\frac{z}{100} \, K_{f} 
\end{equation}
papers from field $f$ among the top $z\%$ papers in the ranking, where $K_f$ is the total number of publications from field $f$ \citep{radicchi2008universality}.
In the following, we denote by $k^{(m)}_f$ the number of publications from field $f$ in the top-$1\%$ of the ranking by metric $m$. We restrict our analysis to $z\% = 1\%$; results for other values of $z$ are available upon request from the authors.

\begin{figure*}[htb]
\centering
\includegraphics[width=\textwidth]{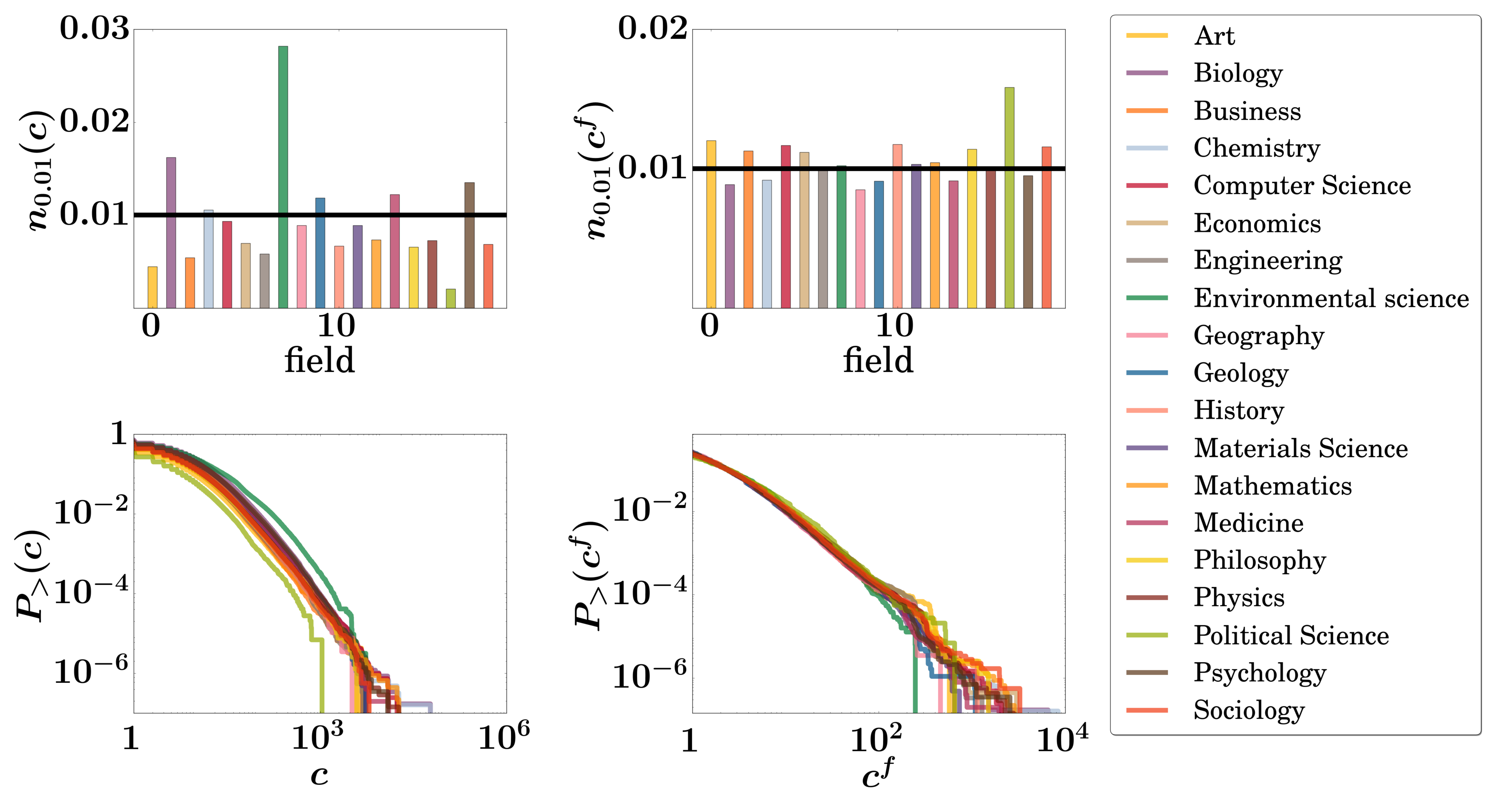}
\caption{Field bias of the analyzed citation-based metrics. Top panels show histograms of the 
fraction of top-1\% publications for each field in the ranking by citation count (left) and relative citation count (right).
The black horizontal line represents the expected value ($0.01$).
Bottom panels show for each field the complementary cumulative distributions for citation count (left) and relative citation count (right). }
\label{fig:rescaled_cit_bad}
\end{figure*}

In the top panels of Fig.~\ref{fig:rescaled_cit_bad}, we illustrate the field bias of citation count, $c$, and relative citation count, $c^f$. 
The presence of strong biases is evident for both metrics because there are fields whose ratio $k^{(m)}_f/K_{f}$ is far away from the expected value $0.01$. 
In particular, Environmental Science is extremely  over-represented in the top of the ranking by citation count.
We argue that this bias comes from the fact that publications from this field have a mean citation count almost twice as big compared to publications belonging to other fields (see Table~\ref{tab:fields}).
For relative citation count, we find a better agreement with what we would expect from an unbiased indicator.
However, relatively large deviations are still evident, especially for the field of Political Science.

In the bottom panels of Fig.~\ref{fig:rescaled_cit_bad}, we report the distributions of $c$ and $c^f$ for each field.
These panels show that the bias by field is not limited to the top $1\%$ papers in the ranking, but it arises from systematic differences between the score distributions across different fields. 
For example, when looking at the distribution of $c$, papers in the field of Political Science have consistently smaller probability to have more than one citation compared to other fields.
For a detailed discussion about the bias of the ranking by $c^f$, we refer to Appendix \ref{appendix:radicchi}.

\begin{figure*}[htb]
\centering
\includegraphics[width=\textwidth]{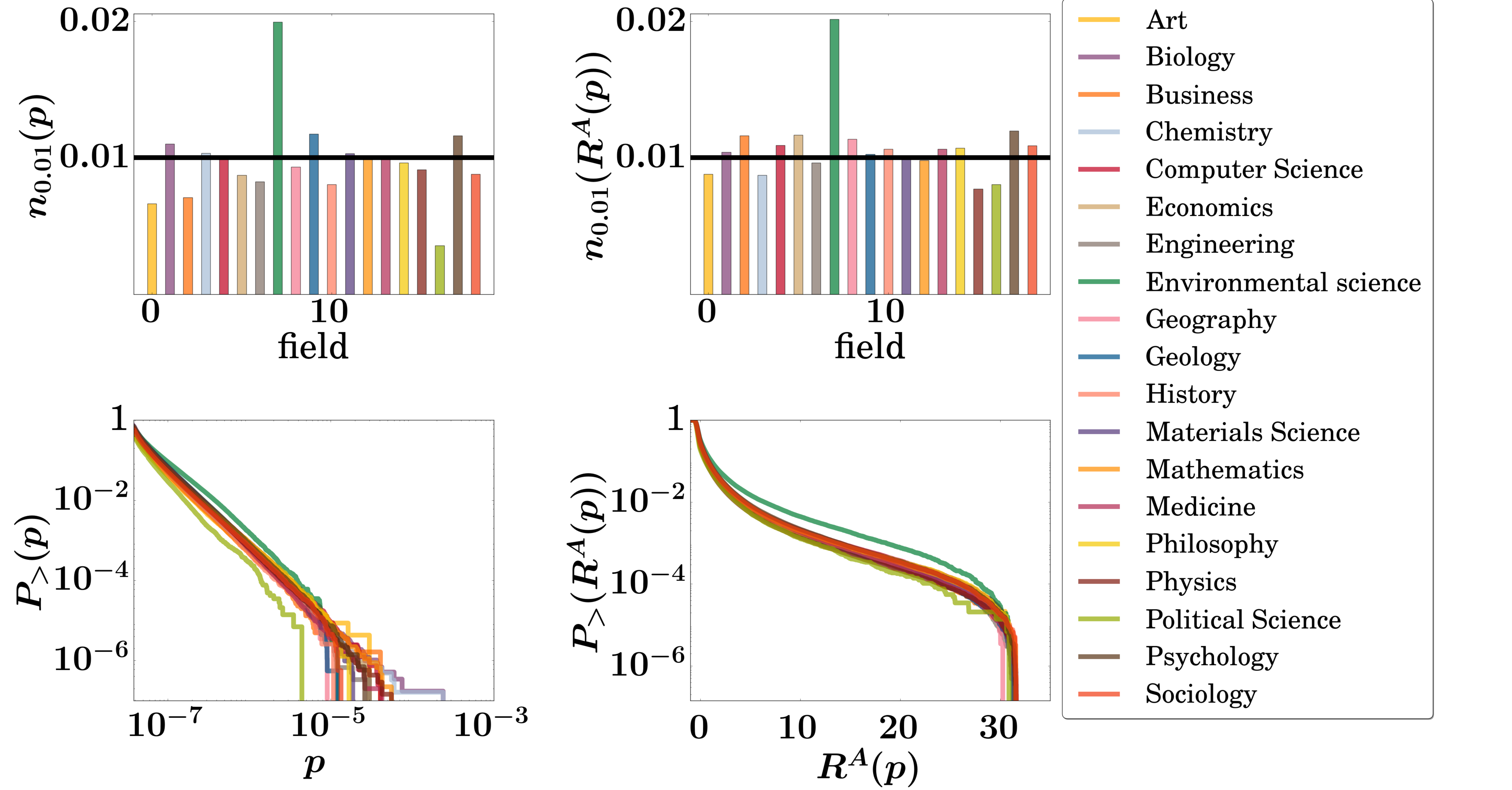}
\caption{Field bias of the analyzed measures based on PageRank. 
Top panels show histograms of the fraction of top-1\% publications for each field in the ranking by PageRank (left) and age-rescaled PageRank (right). 
The black horizontal line represents the expected value ($0.01$).
Bottom panels show for each field the complementary cumulative distributions for  PageRank (left) and age-rescaled PageRank (right). }
\label{fig:rescaled_pg_bad}
\end{figure*}

Fig.~\ref{fig:rescaled_pg_bad} reports the same analysis for PageRank scores, $p$, and age-rescaled PageRank scores, $R^{A}(p)$.
This figure provides the first study of the dependence of PageRank score on academic field.
The top panels of Fig.~\ref{fig:rescaled_pg_bad} show that the top positions of both rankings are biased by field, and both rankings overestimate the impact of publications in the field of Environmental Science. 
Again, we argue that this happens because the mean indegree of publications from Environmental Science is approximately twice as big compared to publications that belong to other fields (see Table ~\ref{tab:fields}).
From the bottom-left panel of Fig.~\ref{fig:rescaled_pg_bad}, we find that the full distribution of scores of Page Rank have a similar shape, but different broadness. 
These differences are slightly smaller for
the age-rescaled PageRank, with the exception of the field of Environmental Science (see bottom right panel of Fig.~\ref{fig:rescaled_pg_bad}). 

\section{Defining new age- and field- normalized metrics}
\label{sec:def-new-metrics}
In this section, we introduce two novel indicators of paper impact: the age- and field-rescaled citation count, $R^{AF}(c)$, and the age- and field-rescaled PageRank, $R^{AF}(p)$. 
The two indicators, $R^{AF}(c)$ and $R^{AF}(p)$, are obtained from citation count $c$ and PageRank score $p$, respectively, through a rescaling procedure.
This procedure is based on the $z$-score and is aimed at suppressing age and field bias.
It is conceptually close to the one presented by \cite{zhang2014comparison}, and is an extension of those introduced by \cite{newman2009first} and \cite{mariani2016identification}.

\subsection{Age- and field-rescaled citation count, $R^{AF}(c)$}

To calculate the age- and field-rescaled citation count $R^{AF}_i(c)$ of a paper $i$ belonging to a field $f$, we first compute the mean $\mu^{AF}_{i}(c)$ and the standard deviation $\sigma^{AF}_{i}(c)$ of the citation count of papers of the \emph{same field} and of \emph{similar age} as paper $i$. 
In particular, $\mu^{AF}_{i}(c)$ and $\sigma^{AF}_{i}(c)$ are computed over the papers that belong to the same field $f$ as paper $i$ and that are among the $\Delta_c$ closest papers to $i$ as measured by the distance $|i-j|$ between their rank by age.
Then, the age- and field-rescaled citation count score $R^{AF}_i(c)$ is defined as 
\begin{equation}
R^{AF}_i(c) = \frac{c_i-\mu^{AF}_i(c)}{\sigma^{AF}_i(c)}.
\label{rc2}
\end{equation}
The averaging window size $\Delta_c$ is a parameter of the method, which we set to  $\Delta_c=1000$.

Differently from \cite{zhang2014comparison}, 
for the computation of the $z$-score, we use temporal windows with the same number of publications, which in general corresponds to real-time intervals of different duration.
This choice is supported by recent findings (\cite{parolo2015attention}) that indicate that in citation networks, time is better defined by number of publications than by real time. 
Our analysis (not shown) confirms that when computation temporal windows are of a fixed temporal length, the bias removal is inferior to that achieved with temporal windows containing a fixed number of publications.

Differently from the relative citation count $c^{f}$, $R^{AF}(c)$ is expected to have not only uniform mean value across different publication dates and fields, but also uniform standard deviation. 
This should lead to a more balanced ranking of the papers. 
We show in the following that this is indeed the case.

\subsection{Age- and field-rescaled Page Rank, $R^{AF}(p)$}
Previous works have shown that PageRank is biased towards old papers in scientific citation networks \citep{chen2007finding, maslov2008promise, mariani2015ranking}.
Moreover, we have shown in Sec.~\ref{sec:short-ex-metric} that PageRank score $p$ is biased by scientific domain.
To simultaneously suppress these two biases, we propose the age- and field-rescaled PageRank score $R^{AF}(p)$.
$R^{AF}(p)$ is defined similarly as $R^{AF}(c)$: we compute the mean value $\mu^{AF}_{i}(p)$ and the standard deviation $\sigma^{AF}_{i}(p)$ of the PageRank scores of the papers that belong to the same field as paper $i$ and that are among the $\Delta_p$ closest papers to $i$ as measured by the distance $|i-j|$ between their rank by age.
The age- and field-rescaled PageRank score is then defined as
\begin{equation}
R^{AF}_{i}(p) = \frac{p_i-\mu^{AF}_i(p)}{\sigma^{AF}_i(p)}.
\label{rpr2}
\end{equation}
In the following, we set $\Delta_p=1000$.

\begin{figure*}[htb]
\centering
\includegraphics[width=\textwidth]{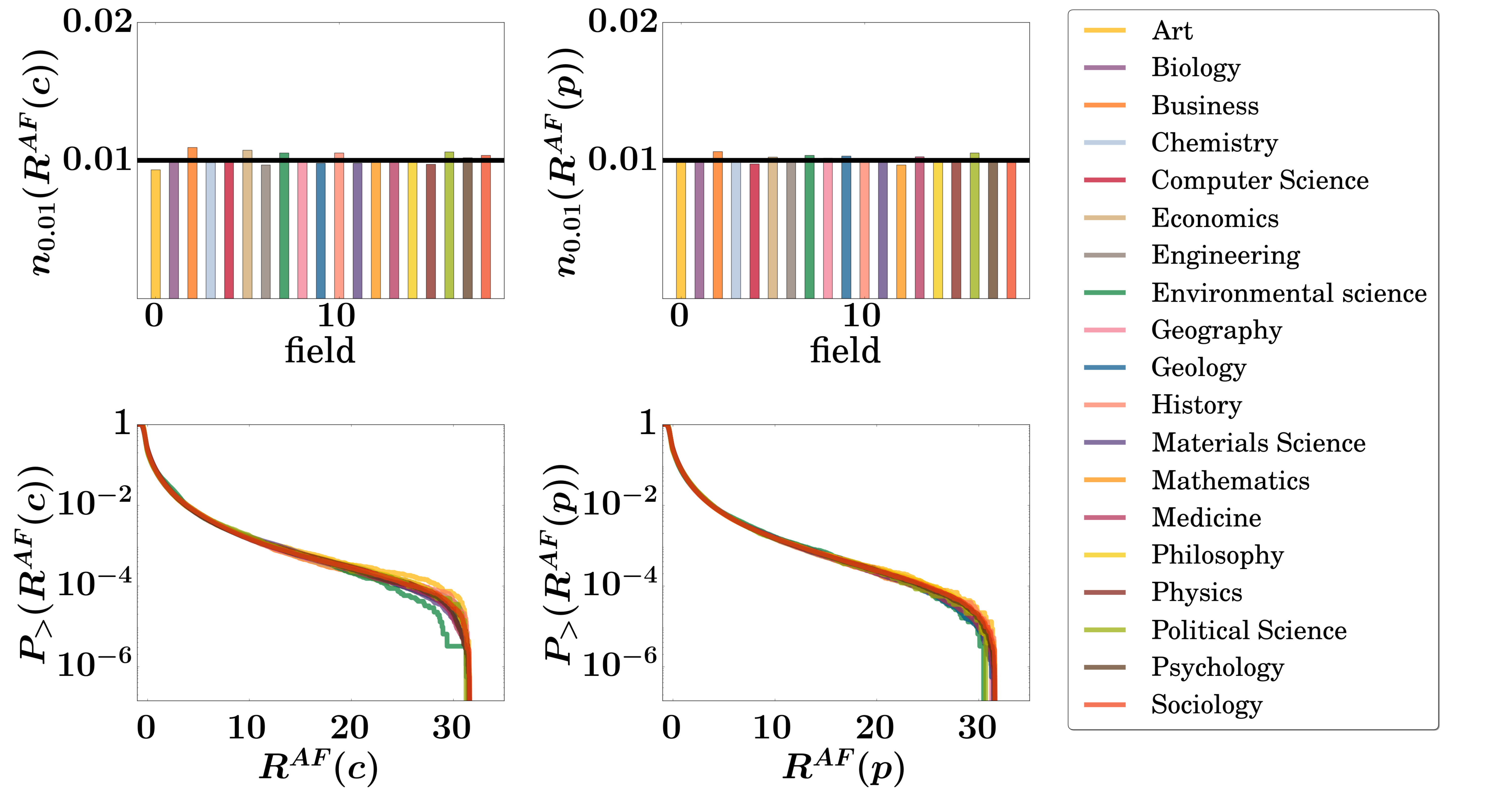}
\caption{Field balance of the analyzed citation-count and PageRank-based metrics. 
Top panels show histograms of the number of top-1\% publications for
each field in the ranking by age- and field-rescaled citation count (left), and age- and field-rescaled PageRank (right panel).
The black horizontal line represents the expected value ($0.01$).
Bottom panels show for each field the complementary cumulative distributions for
age- and field-rescaled citation count (left), and age- and field-rescaled PageRank (right).
}
\label{fig:rescaled_good}
\end{figure*}

In the top panels of Fig.~\ref{fig:rescaled_good}, we show that in the top-1\% of the rankings by $R^{AF}(p)$ and $R^{AF}(c)$ each field appears well represented. 
In fact, the deviations from the expected value are very small especially if compared to the deviations of the other rankings (see top panels in Figs.~\ref{fig:rescaled_cit_bad} and \ref{fig:rescaled_pg_bad}).
In the bottom panels of Fig.~\ref{fig:rescaled_good}, we report that the the full score distributions for papers from different fields collapse extremely well top of each other thanks to the rescaling procedure.

\section{Quantifying rankings' biases by field and age}
\label{sec:quant-rank-bias-by-field-and-age}
We begin this section by introducing a new methodology to assess a ranking's bias based on the Mahalanobis distance (paragraph \ref{sec:mahala}). 
Then, we use this to quantify the bias by field (paragraph \ref{sec:field_bias2}) and the bias by age and field (paragraph \ref{sec:field_age_bias}).

\subsection{A general framework to assess ranking biases based on the Mahalanobis distance}
\label{sec:mahala}

While Figs.~\ref{fig:rescaled_cit_bad} and~\ref{fig:rescaled_pg_bad} illustrate the substantial field bias of the existing metrics, the bias is much weaker (if any) for the new age- and field-rescaled metrics in Figure~\ref{fig:rescaled_good}.
Now we quantify this improvement by extending the statistical tests of bias suppression presented by \cite{radicchi2008universality,waltman2012universality,radicchi2012testing,mariani2016identification}. 
Similarly to these works, we assume that a ranking is unbiased if its properties are consistent with those of an unbiased selection process.

\subsubsection{Assessing the bias by field}

Let us first analyze the problem of assessing the bias by field. 
Consider an urn which contains $N$ marbles, each of them corresponding to one of the publications present in our dataset.
An \emph{unbiased selection process} then corresponds to sampling from this urn at random without replacement a fixed number $n\!=\!\lfloor N\times 0.01 \rfloor$ of publications.
From the extracted sample, we count the number of publications that belong to each field $f$, $k_f$, and record these numbers in the vector $\vec{k}=(k_1,... ,k_F)^T$; here $F$ denotes the number of fields. 
The probability to observe a certain vector, $\vec{k}$, is given by the \emph{multivariate hypergeomentric distribution} (MHD)
\begin{equation}\label{eq:MHD}
P\left(\vec{k}\right)=\frac{\prod_f^F \binom{K_f}{k_f}}{\binom{N}{n}}
\end{equation}
where $K_f$ is the total number of publications in field $f$.
Following this selection process, among the $n$ extracted publications, the expected number of publications for field $f$ is $\mu_f\!=\!n\,K_f/N$. 

Assume that the actual ranking by a given metric $m$ features $k^{(m)}_f$ publications from field $f$ in the top $1\%$ of its ranking. 
In general, the observed $k^{(m)}_f$ deviates from its expected value $\mu_f$.
A simple approach to quantify this deviation would consist in computing the $z$-score, defined as $z^{(m)}_f:=(k^{(m)}_f-\mu_f)/\sigma_f$, where $\sigma_f$ is the expected standard deviation for field $f$ according to the MHD specified by Eq.~\eqref{eq:MHD}.
There are however two shortcomings of the $z$-score. 
First, the $z$-score only gives partial information for a MHD -- how far from the expected values we are in units of standard deviations -- but it does not provide information on how statistically significant the deviations are.
Second, to quantify the overall bias of a given indicator $m$, we would need to aggregate the $z$-scores from the different fields. For example, we could take the average $z$-score, but this would neglect the correlation between the different fields coming from the constraint $n=\sum_f k^{(m)}_f$.

\begin{figure*}[htb]
\centering
\includegraphics[clip, trim=20mm 20mm 0mm 18mm, width=\textwidth ]{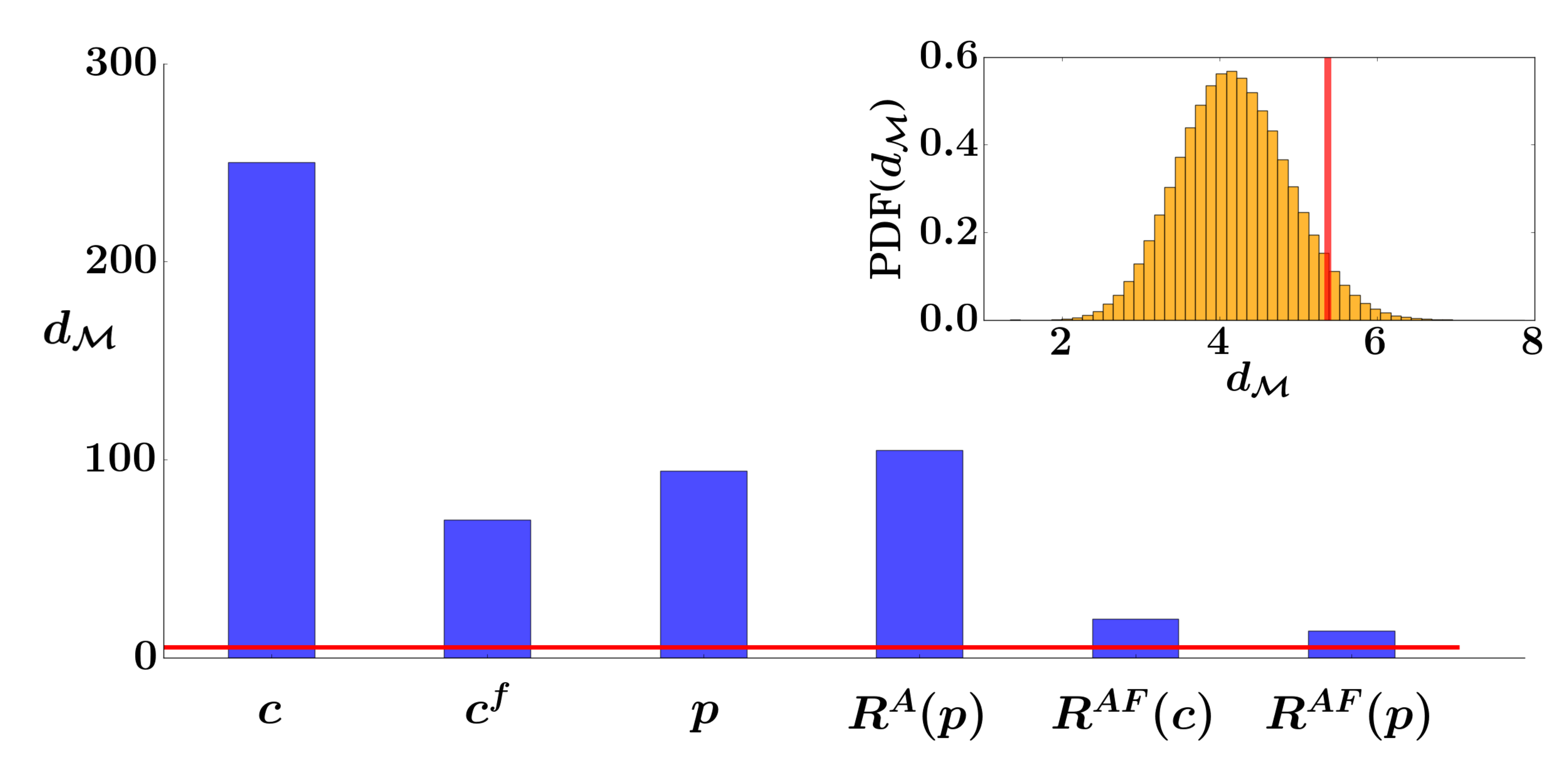}
\caption{Mahalanobis distance, $d_{\mathcal{M}}$, for the analyzed indicators when considering the $19$ main fields. 
From left to right: citation count, relative citation count, PageRank, age-rescaled PageRank, 
age- and field-rescaled citation count and age- and field-rescaled PageRank. 
The horizontal red line represents the upper bound of the $95\%$ confidence interval obtained from the simulations. 
In the insets, we report the distribution of $d_{\mathcal{M}}$ coming from $1\,000\,000$ simulations of the unbiased sampling process. 
Again, the red line represents the upper bound of the $95\%$ confidence interval.
}
\label{fig:MD_field}
\end{figure*}

To overcome these two problems, we follow a different approach. 
We first run various numerical simulations that reproduce the unbiased selection process.
These simulations produce a set of ranking vectors which are distributed according to Eq.~\eqref{eq:MHD} around the vector of expected values, $\vec{\mu}\!=\!(\mu_1,...,\mu_m)$.
Differently from~\citep{radicchi2012testing}, we do not estimate the confidence interval for the different fields separately. 
We calculate instead the Mahalanobis distance ($d_{\mathcal{M}}$,~\cite{Mahalanobis1936distance} and Appendix \ref{app:MD}) for each simulated vector from $\vec{\mu}$, and construct the distribution of $d_{\mathcal{M}}$'s obtained by the simulated unbiased selection process.
The inset of the left panel of Fig.~\ref{fig:MD_field} reports the distribution of the $d_{\mathcal{M}}$ for $1\,000\,000$ simulations. 
The distribution is centered around its mean value of $4.18$ and 
the upper bound for the $95\%$ confidence level
is around $5.37$. \footnote{A curiosity for the reader. 
Here, the average of the square of the $d_{\mathcal{M}}$ for the unbiased sampling process is extremely close to the number of degrees of freedom of our problem. 
This stems from the fact that the MHD defined by Eq.~\eqref{eq:MHD} converges to a Multivariate Gaussian Distribution (MGD) as we increase the number of publications $N$ while keeping $n/N$ fixed and small. 
Our dataset is large enough for this approximation to be accurate. The $d_{\mathcal{M}}^2$ of a MGD is distributed as a $\chi^2$ variable with average equal to the number of degrees of freedom, i.e. 18 since we have 19 distinct fields and one constraint.}
For an \emph{ideal unbiased ranking}, we would expect its $d_{\mathcal{M}}$ to fall into the $95\%$ confidence interval of the distribution of the $d_{\mathcal{M}}$ obtained from the simulated unbiased sampling process. 

\subsubsection{Assessing the bias by age and field}
The methodology presented above is easily generalized to simultaneously assess a ranking's bias by age and field.

To add the temporal dimension to the bias assessment procedure, we split the publications into $T$ equally-sized age groups, and repeat the above analysis by using $F\times T$ different categories of publications. 
In Sec.~\ref{sec:field_age_bias}, we set $F=19$ representing the number of fields and $T=40$ as in \citep{mariani2016identification}, and thus we obtain $760$ age-field groups of different sizes. 

\subsection{Results on the bias by field}
\label{sec:field_bias2}

The rankings produced by different metrics differ greatly by their $d_{\mathcal{M}}$ (see Fig.~\ref{fig:MD_field}). 
As expected, the metrics that are not field-rescaled ($c$, $R^{A}(p)$, and $p$) are far from being unbiased. 
At the same time, relative citation count that is rescaled by field performs only slightly better than PageRank which is ignorant of any field information. 
The best results by a wide margin are achieved by our metrics, $R^{AF}(c)$ and $R^{AF}(p)$, obtained using the new rescaling. 
Nevertheless, both these metrics fail to meet the 95\% upper bound achieved by simulated unbiased rankings. 
As disappointing as it may seem, this finding is not entirely surprising as the proposed rescaling procedures focus on equalizing the first two moments of the respective quantities ($c$ and $p$) whereas the quantities' distributions can differ also by higher moments. 

To understand which field contributes the most to the resulting $d_{\mathcal{M}}$ values, we derived an alternative analytic expression for the $d_{\mathcal{M}}$
\begin{equation}\label{eq:contr_md}
d_{\mathcal{M}}(\vec{k},\vec{\mu})^2 = \sum_i^F z_i^2 \left(1 - \frac{k_i}{N}\right)
\end{equation} 
where we omit the metric superscript $(m)$ from the notation for $z_i$ and $\vec{k}$ for simplicity.
We have proven this formula analytically for $F=3,4,5,6$, and we have numerically tested it for $F=19$ 
and $760$ (see Appendix \ref{app:MD}); it remains open to prove it in arbitrary dimensions.

\begin{table*}
\centering
\begin{tabular}{rrrrrrr}
  \hline
  Field & \bf $c$  & \bf $c^f$ & \bf $p$  & \bf $R^{A}(p)$ & \bf $R^{AF}(c)$  & \bf $R^{AF}(p)$\\
  \hline
  Art & 1.15 & 1.95 & 3.01 &0.31 & 2.84 &0.09 \\
  Biology &\textbf{36.46} &15.81 &6.74 & 0.87 & 0.12 & 0.16\\
  Business & 2.06 & 2.08 & 5.95 & 1.51 & 14.56 & 13.50 \\
  Chemistry & 0.34 & 8.44 & 0.85 & 9.28 & 4.23 & 0.00 \\
  Computer Science & 0.29 & \textbf{23.86} & 0.02 & 3.09 & 0.12 &13.50 \\
  Economics & 3.34 & 6.51 & 4.20 & 5.77 & \textbf{32.32 }& 7.93 \\
  Engineering & 8.36 & 0.09 & 10.48 & 0.35 & 8.36 &0.03 \\
  Environmental Science & 16.82 & 0.04 & \textbf{35.67} & 29.89& 2.45 &2.23 \\
  Geography & 0.05 & 1.34& 0.15 &0.50 & 0.15 &0.51 \\
  Geology & 1.02 & 3.04 & 6.42 &0.13 & 2.10 &10.06 \\
  History & 0.69 & 2.48 & 1.67 &0.15 & 3.12 &0.52 \\
  Materials Science & 0.39 & 0.43 & 0.23 &0.01 & 2.37 &0.10\\
  Mathematics & 5.17 & 1.94 & 0.10 &0.14 & 0.09 &\textbf{25.34} \\
  Medicine & 4.02 & 7.66 & 0.06 &1.94 & 2.16 &19.56\\
  Philosophy & 1.49 & 3.23 & 0.12 &0.36 & 0.15 &0.07 \\
  Physics & 8.30 & 0.21 & 6.00 &\textbf{33.67}& 14.34&0.01 \\
  Political Science & 1.47 & 10.22 & 6.82 &0.51 & 1.47 &2.44 \\
  Psychology & 5.75 & 1.43 & 8.56 &10.24& 2.93 &1.65 \\
  Sociology & 2.83 & 9.23 & 2.95 & 1.30 &6.11 & 2.30 \\
  \hline
\end{tabular}
\caption{The individual contribution $z_{i}^{2}\,(1-k_{i}/N)$ of each field $i$ to the square $d_{\mathcal{M}}^{2}$ 
of the Mahalanobis distance for the different metrics.}
\label{tab:MD_contribution}
\end{table*}
In Table \ref{tab:MD_contribution}, we report the individual fields' contributions to the $d_{\mathcal{M}}^2$ calculated using Eq.~\ref{eq:contr_md}.
We find that Biology and Computer Science are the fields which give the biggest contributions to the $d_{\mathcal{M}}^2$ for the rankings by citation count and relative citation count, respectively. 
This could not have been detected by looking at the deviations from the expected values.
Indeed, in Fig.~\ref{fig:rescaled_cit_bad} we only see that Environmental Science and Political Science have the largest deviations.
For the novel indicators, approximately one third of the $d_{\mathcal{M}}^2$ of $R^{AF}(c)$ is explained by the field of Economics and approximately one fourth of the $d_{\mathcal{M}}^2$ of $R^{AF}(p)$ is explained by the field of Mathematics. 

In addition, we also find that the $d_{\mathcal{M}}$'s contributions across different fields assume values in a relatively broad range. 
This suggests that findings on rankings' bias by field may strongly depend on which disciplines are included or not in the analysis. We argue that arbitrary choices on which fields to include should be avoided in future research on field-normalization of impact indicators.

To summarize, our bias suppression test allows us not only to estimate the level of bias ($d_{\mathcal{M}}$) of the various metrics, but also to quantify which percentage of the total bias ($d_{\mathcal{M}}^2$) of an indicator is explained by each single field.

\subsection{Results on the bias by age and field}
\label{sec:field_age_bias}

While the analysis of the previous paragraphs focused on the ranking bias by field, in this paragraph we use the $d_{\mathcal{M}}$ to simultaneously assess the bias by age and field of a given ranking.

\begin{figure*}[htb]
\centering
\includegraphics[clip, trim=20mm 20mm 0mm 18mm, width=\textwidth ]{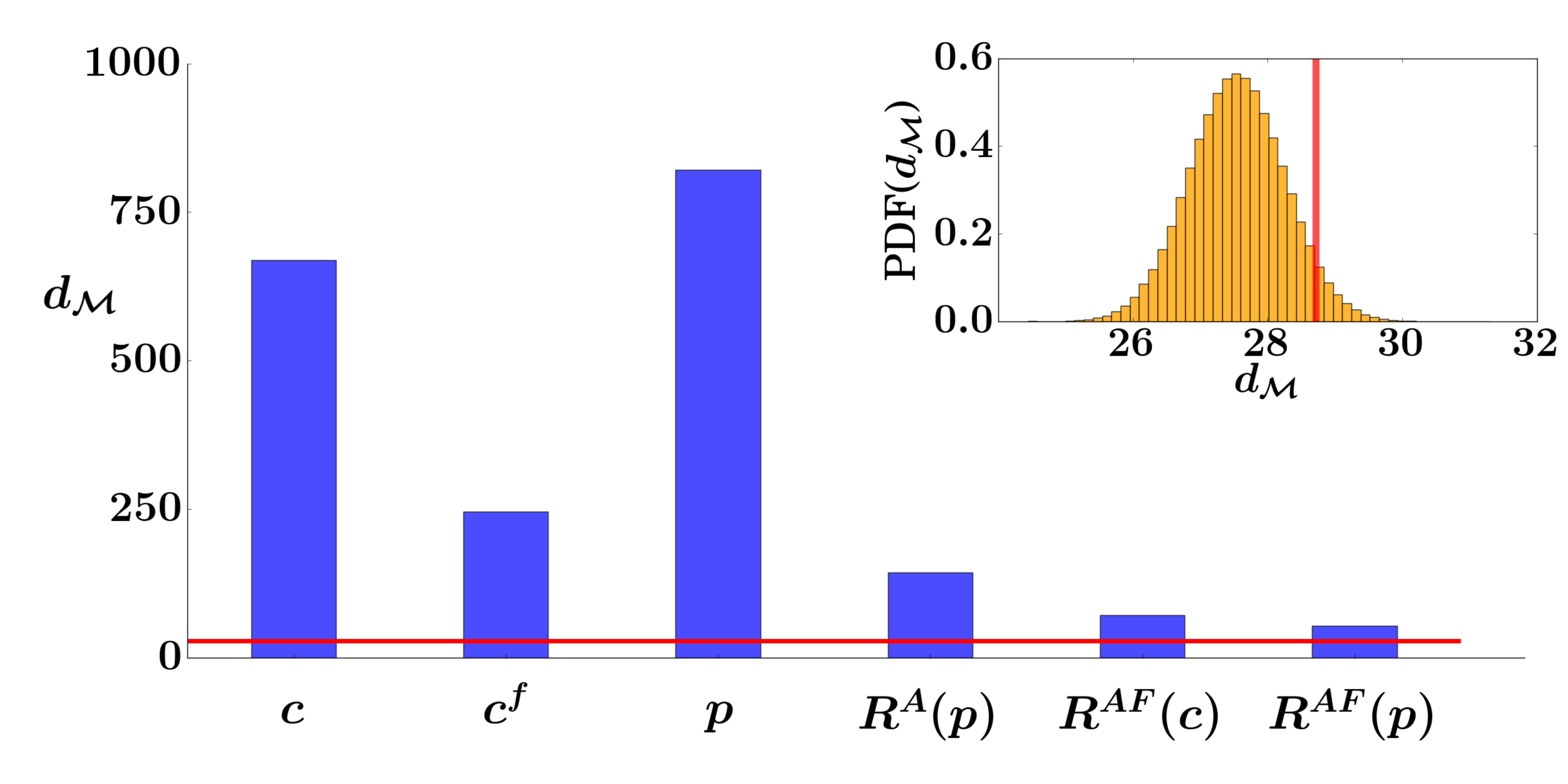}
\caption{Mahalanobis distance, $d_{\mathcal{M}}$, for the analyzed indicators when considering the $760$ age-field groups.
From left to right: citation count, relative citation count, PageRank, age-rescaled PageRank, age- and field-rescaled 
citation count and age- and field-rescaled PageRank. 
The horizontal red line represents the upper bound of the $95\%$ confidence interval obtained from the simulations. 
In the insets, we report the distribution of $d_{\mathcal{M}}$ coming from $1\,000\,000$ simulations of the unbiased sampling process. 
Again, the red line represents the upper bound of the $95\%$ confidence interval.
}
\label{fig:MD_field_2}
\end{figure*}
In Fig.~\ref{fig:MD_field_2}, we show the $d_{\mathcal{M}}$'s for the different indicators and for the 95\% confidence interval for the simulated unbiased selection process using $40\times 19$ age-field types of publications. 
For citation count, PageRank, relative citation count and age-rescaled PageRank we have to reject the hypothesis that the rankings of these indicators are not biased by age and field. 
For the improved indicators, age- and field-rescaled citation count and PageRank, we also have to reject the null hypothesis, even though they are much closer to the  95\% confidence interval. 

It is worth to notice that age-rescaled PageRank, an indicator developed to only remove PageRank's bias by age (\cite{mariani2016identification}), is significantly less biased compared to relative citation count, an indicator specifically designed to simultaneously remove bias by age and field. 

\subsection{Simultaneously visualizing the bias by age and field}

\begin{figure*}[htb]
\centering
\includegraphics[clip, trim=43mm 100mm 45mm 70mm, width=\textwidth ]{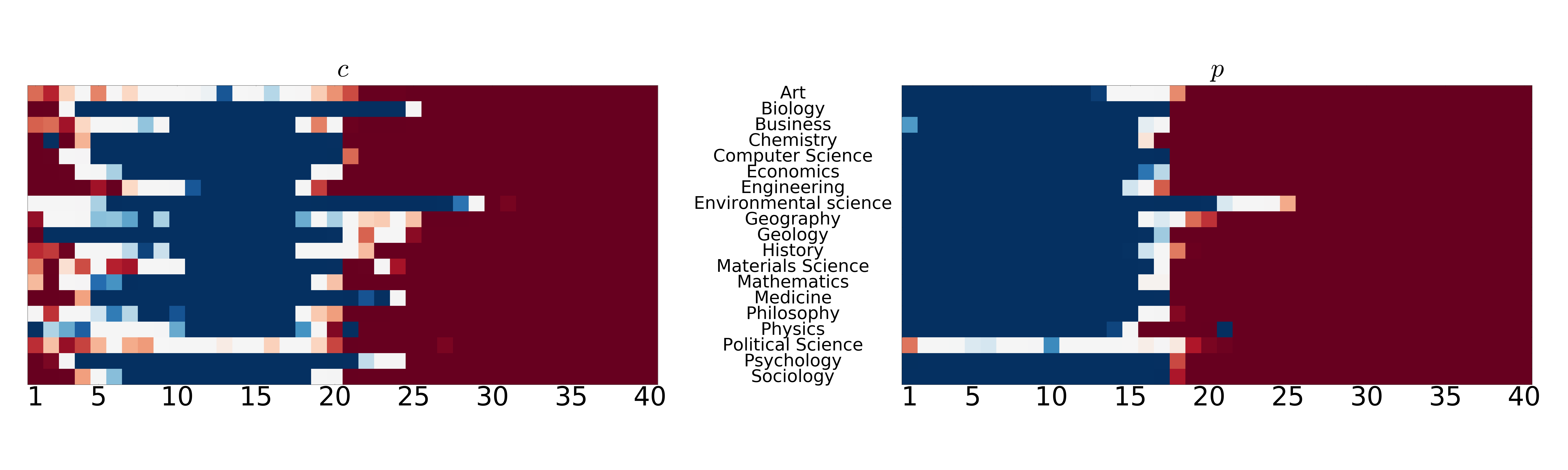}\\
\includegraphics[clip, trim=43mm 100mm 45mm 70mm, width=\textwidth,  ]{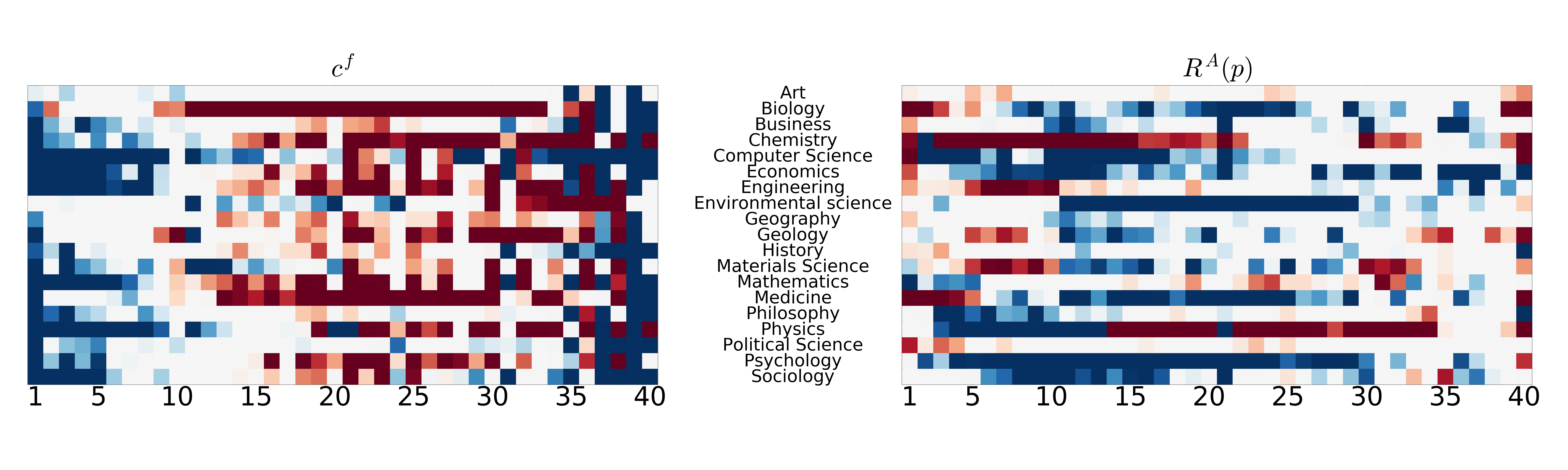}\\
\includegraphics[clip, trim=43mm 0mm 45mm 5mm, width=\textwidth ]{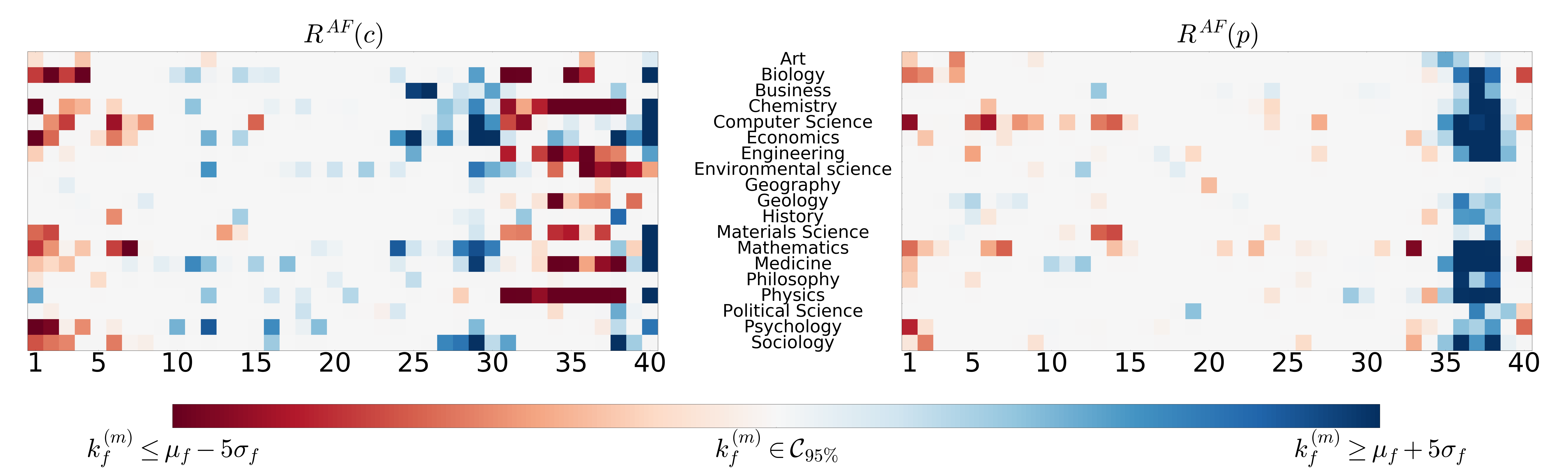}
\caption{Heatmaps showing the bias by field and age of the rankings by the different indicators. 
Each cell represents an age-field group: age groups are represented on the horizontal axis, while fields are represented on the vertical axis.
The color of the cells shows the bias of the indicators with respect to that age-field group. 
White means that the respective age-field group is fairly represented in the top $1\%$ of the ranking by the indicator,
whereas we use a color scale from white to intense red (blue) for age-field groups which are underestimated (overestimated).
}
\label{fig:age_field}
\end{figure*}

To visualize the field and age bias of the rankings by the analyzed indicators, we use 
heatmaps in the age-field group plane (see Fig.~\ref{fig:age_field}).
In these heatmaps, each cell represents a field-age group, and its color indicates the level of bias. 
A white cell indicates that the number of papers in the respective age-field group falls into the 95\% confidence level ($\mathcal{C}_{95\%}$) determined with the simulations. Hence, white means that no bias is detected for that age-field group.
While for representing the bias towards or against a group of papers, we use blue (overestimation) and red (underestimation).
To obtain a range of over/under-estimation, the brightness of the colors ranges from white (no bias) to intense blue/red.
The most intense colors indicate that the number of papers from that age-field group is $5$ standard deviation smaller/bigger than the expected value.

The top panel of Fig.~\ref{fig:age_field} shows that, independently of field, citation count and PageRank systematically over-represent old papers and under-represent recent papers.
This is in agreement with the findings of several other works \citep{chen2007finding,newman2009first,mariani2015ranking,mariani2016identification}.
The only exception is Political Science which is usually underestimated independently of paper age. 
We argue that this happens because this is the smallest field in the data set, and it has become an academic discipline by itself much later compared to most of the other fields. 
Also, the oldest papers in most fields are under-represented by citation count, which reflects the change of citation practices over time.

The middle panels of Fig.~\ref{fig:age_field} show that the relative citation count and age-rescaled PageRank suppress large part of the biases of the original metrics, yet specific fields are consistently overestimated or underestimated. 
For example, both age-rescaled PageRank and relative citation count under-represent papers belonging to the field of Chemistry. 
A peculiarity of relative citation count is that it over-represents both the oldest as well as the most recent papers at the cost of the other papers.

The bottom panels of Fig.~\ref{fig:age_field} show the heatmaps for the new indicators age- and field- rescaled citation count $R^{AF}(c)$ and PageRank $R^{AF}(p)$. 
We find that the respective rankings are much less biased towards specific fields compared to all the other analyzed measures.
However, there are two patterns: for $R^{AF}(c)$  recent publications tend to be underestimated for some fields, whereas for $R^{AF}(p)$ recent publication tend to be overestimated for almost all fields.
These rather systematic patterns must have their roots in changes of the citation and PageRank score distributions with time.
Since our rescaling procedure was fixing the first two moments of these distributions, the observed patterns come from differences in higher moments.
Thus, the distributions of $R^{AF}(c)$ and $R^{AF}(p)$ are aligned only partially for papers of different age.

\section{Discussion and conclusion}

To summarize, in this paper we introduced two new indicators of paper impact, rescaled citation count $R^{AF}(c)$ and rescaled PageRank $R^{AF}(p)$, and we analyzed a large citation network from the Microsoft Academic Graph to show that the rankings of papers by these indicators are much less biased by age and field than the rankings by other existing indicators.
In particular, the ranking by $R^{AF}(p)$ is approximately three times less biased compared to the least biased existing metrics, relative citation count and age-rescaled PageRank.
The level of bias of the rankings has been quantified with a new statistical framework based on the Mahalanobis distance. 
This framework allowed us to simultaneously quantify the age and field biases of the analyzed rankings, and to determine which groups of papers give the largest contributions to the observed bias.

The contribution of our results to the debate on the validity of field-normalization procedures is threefold. 
First, our findings are in agreement with the conclusions of \cite{albarran2011skewness} and \cite{waltman2012universality} which argued that the relative citation count introduced by \cite{radicchi2008universality} can be insufficient to effectively remove citation count's bias by age and field.
Second, we show the importance of testing indicators using an accurate statistical procedure, such the one introduced in this paper. Indeed, for the least-biased indicators analyzed, $R^{AF}(c)$ and $R^{AF}(p)$, no clear indication of bias is found at first glance. However, when using the statistical test based on the Mahalanobis distance, we find a significant discrepancy between their rankings and those coming from unbiased sampling process. We argue that including higher-order momenta (such as the skewness) in the rescaling procedure can be an efficient way to further reduce the rankings' level of bias.
Third, by deriving an explicit formula to calculate the contribution of each field to the bias of a ranking, we find that the these contributions assume a broad range of values.
We obtain similar findings also for the contributions to the age-field bias.
This means that the level of bias of rankings depends heavily on which years or fields are included in the analysis.
For this reason, in future research on age and field normalization of indicators, it is essential to clearly motivate which years and fields are included in the analysis, avoiding arbitrary or uncritical decisions.

To address the bias by age and field of ranking of papers, we have first divided the papers in groups with similar age and from the same field. 
Then, we considered only the sizes of these groups to define an unbiased selection process from which we obtained a statistical null model for an unbiased ranking.
In principle, many other information can be included into the null model to correct for other effects. 
For example, including information about the co-authorship network would permit to correct for the effect of this network on the growth and structure of the citation network \citep{sarigol2014predicting, Schweitzer2017}.
In this way, we would gain a better understanding of how the social dimension of science contributes to the field and age biases of impact indicators.

We emphasize that removing the biases addressed in this paper and those that come from social aspects is of primary importance not only for scholarly publication databases, but also for several other information systems, such as the WWW or online social networks \citep{scholtes2014social}.
As a matter of fact, every day scholars and on-line users explore available knowledge using recommender systems based on ranking algorithms.
This challenges us to design more sophisticated filtering and ranking procedures to avoid biases that can systematically hide relevant contents or only show information too similar to what the users already know.

While we have analyzed in detail the presence of age and field bias in the ranking, it still remains to evaluate the actual ranking performance of the newly proposed indicators in artificial data~\citep{medo2016model} or in real data where the ground truth is provided by some external source~\citep{dunaiski2016evaluating, mariani2016identification}.
Another important issue is the comparison between metrics based on citation count and metrics that take the whole citation network into account to determine papers' score.
Our analysis (see Appendix ~\ref{sec:comparing}) shows that the rankings by the least-biased indicators $R^{AF}(c)$ (citation-based) and $R^{AF}(p)$ (network-based) are positively correlated, still substantially different.
Can we use the extra-information provided by the network to enhance our ability to identify highly-significant publications?
Our intuition and the results presented by \cite{chen2007finding} and by \cite{mariani2016identification} for specific research fields suggest that this is the case. Yet, we need additional analysis to validate this conjecture in larger datasets such as the one analyzed in this article.

To conclude, by reducing the age and field biases from indicators of scientific impact and by extending the existing statistical tests for biases, we contribute to the challenge of quantifying and suppressing biases of rankings in information systems.

\section*{Acknowledgments}
We thank Ingo Scholtes, Frank Schweitzer and Yi-Cheng Zhang for suggestions which improved the manuscript. In addition, we also thank Emre Sarig\"ol for his help in pre-processing the data, and Elias Bauman for his important contribution to the optimization of the C++ code that we used for the network analysis.
GV acknowledges support from the Swiss State Secretariat for Education, Research and Innovation (SERI), Grant No. C14.0036 as well as from EU COST Action TD1210 KNOWeSCAPE. MSM acknowledges support from the Swiss National Science Foundation Grant No. 200020-143272.

\bibliographystyle{unsrtnat}

\clearpage

\appendix

\section{KDD Cup data}
\label{app:kdd}
\subsection{Data Source}
In this work, we analyzed the dump of the \textit{Microsoft Academic Graph} (MAG) released for the KDD Cup 2016 \citep{sinha2015overview}, a competition linked to a prestigious computer science conference on knowledge discovery and data mining (KDD). 
Among the primary interests of the community organizing the KDD Cup, there are the technical challenges related to web-scale data collection and aggregation.
For this reason, the released data for the KDD Cup 2016 went through only basic processing\footnote{\url{https://kddcup2016.azurewebsites.net/Data}}.
Each publication in the dataset is endowed with its unique identifier, \emph{paperid}, \emph{publication date}, \emph{references} to other publications, and its \emph{field of study}.
\subsection{Data pre-processing}
When analyzing the data, we do not distinguish the publications by their type (paper, review, book, etc.).
Further, we also do not differentiate between different types of journals and take into account all of them: for example, we do not distinguish between a citation coming (or going) from (or to) a letter or a book. 
We argue that it is important to keep various types of journals and publications because different fields adopt not only different citation norms, but also different ways to communicate their results.
For example, computer science researchers commonly publish results in conference proceedings, while physics authors tend to prefer articles or letters.
At the same time, we are aware that different types of publications might have different citation characteristics.
However, good indicators should ideally be able to account for heterogeneities among publications and citation norms across different communities and produce unbiased rankings without the need for arbitrary choices about which types of articles to include in the analysis.
In addition, as \cite{waltman2012universality}, we do not exclude publications which do not receive citations.

As mentioned in Sec.~\ref{sec:data}, the MAG has a field classification scheme with 4 hierarchical levels.
The field assignment is based on an internal algorithm that uses a machine learning approach \citep{sinha2015overview}. 
In our work, similarly to \cite{radicchi2008universality}, we are only interested in impact metric normalization at the most coarse-grained level. 
To this aim, in our analysis, we focus only on the $19$ main fields as listed in Table~\ref{tab:fields}. 
Discussing the possible limitations of the classification approach by MAS and the dependence of our results on the adopted classification scheme is a relevant subject for future research but goes beyond the scope of this manuscript.
We only included in the analysis papers for which the following information are available: (1) unique identifier (ID); (2) complete publication date (yyyy/mm/dd) crucial for the temporal rescaling procedure that is explained in the following; (3) DOI or journal-id, in order to be able to retrieve the publication; (4) assignment to at least one of the main $19$ fields. 
We discard from our analysis publications for which one or more of these four properties are missing. 

With this filtering procedure, we obtain $N=18\,193\,082$ unique publications and $E=109\,719\,182$ citations.

\begin{table}
\centering
\begin{tabular}{rrr}
\hline
Field & Publication count  & Mean citation count\\
\hline
Art &233\,251 &3.90\\
Biology &5\,847\,554  &9.67\\
Business &613\,827 &4.81\\
Chemistry &6\,204\,531 &7.28\\
Computer Science & 4\,080\,636 &6.13\\
Economics &2\,252\,921 &5.56\\
Engineering &3\,011\,763 &5.10\\
Environmental Science &315\,465 &12.63\\
Geography &288\,338 &6.92\\
Geology &1\,825\,707 &7.88\\
History &390\,144 &5.53\\
Materials Science &2\,063\,474 &6.12\\
Mathematics &4\,551\,453 &5.87\\
Medicine &5\,061\,990 &7.90\\
Philosophy &787\,649 &5.05\\
Physics &6\,976\,644 &5.55\\
Political Science &144\,473 &2.51\\
Psychology &2\,861\,813 &8.23\\
Sociology &1\,784\,695 &5.39\\
\hline
Total & 49\,296\,327 & -- $ \, \, \, $ \\
\hline
Total (no multiple) & 18\,193\,082 & 6.42\\
\hline
\end{tabular}
\caption{{\bf Main Fields} The $19$ parent main fields identified by Microsoft Academic Graph with their number of publications and average citations. The second to last row reports the total number of publications considering multiple times publications that belong to more than one fields, whereas the last row reports the total number of unique publications and the average citation count.}
\label{tab:fields}
\end{table}

\subsection{Data basic statistics}
\begin{figure}
\centering
\includegraphics[scale=0.4]{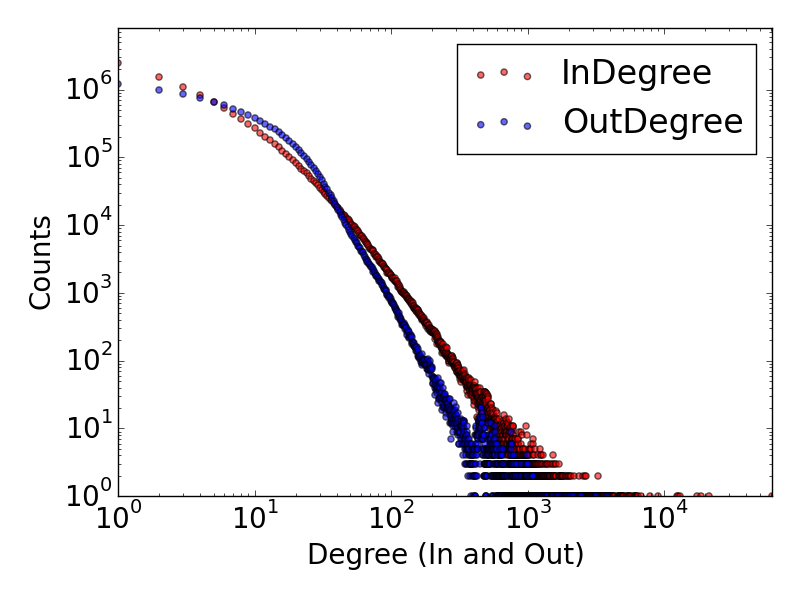}
\caption{In- and out-degree distribution for the publications present in our dataset after preprocessing the data.}
\label{distribution}
\end{figure}

We first observe that the distributions of both incoming and outgoing edges are broad (see Fig.~\ref{distribution}).
Both these distribution in fact have long and heavy tails. 
In addition, we also observe strong variations of the mean citation count across fields (see Table \ref{tab:fields} for details).
For example, publications belonging to the field of Environmental Science and of Political Science have mean citation count two times bigger and smaller, respectively, compared to the average citation count across all fields.
In agreement with the findings by \cite{schubert1986relative, vinkler1986evaluation, radicchi2008universality}, the strong variety in mean citation count per publication for the various fields confirms that the corresponding communities exhibit different citation behavior, which calls for field normalization procedures.

\subsection{Data quality}
Here, we address the following question: to which extent is the KDD dump of the MAG in accordance with the most updated online version of the MAG\footnote{We have performed this analysis during February 2017.}? 
For this comparison, we divided the $49\,296\,327$ analyzed paper-field pairs (one paper-field pair is composed of a paper and the fields that paper belongs to) into $760$ age and field groups, as described in Sec. \ref{sec:quant-rank-bias-by-field-and-age}. 
From each group, we randomly choose the $0.1\%$ of papers. 
Following this procedure, we obtained a representative sample with respect to field and age composed of approximately $50\,000$ papers.

First, we have matched the paper ID in hexadecimal format present in the data released for the KDD Cup to the paper ID in int64 format present in the on-line version of the MAG. 
For $50$ papers, we manually verified that the paper IDs were exactly the same in the two datasets, albeit represented in different formats.
Then, using the Academic Knowledge Api\footnote{\url{https://www.microsoft.com/cognitive-services/en-us/academic-knowledge-api}}, we have downloaded the number of citations for each sampled paper.

For the $2.3\%$ of the sampled papers, we were not able to find their corresponding paper in the online MAG.
For $50$ unmatched papers, we found out that these were papers with duplicates in the KDD Cup data, i.e. these papers are present in the KDD data with two distinct IDs, and only one of the two IDs is present in the MAG.

\begin{figure}[htb]
\centering
\includegraphics[width=0.55\textwidth]{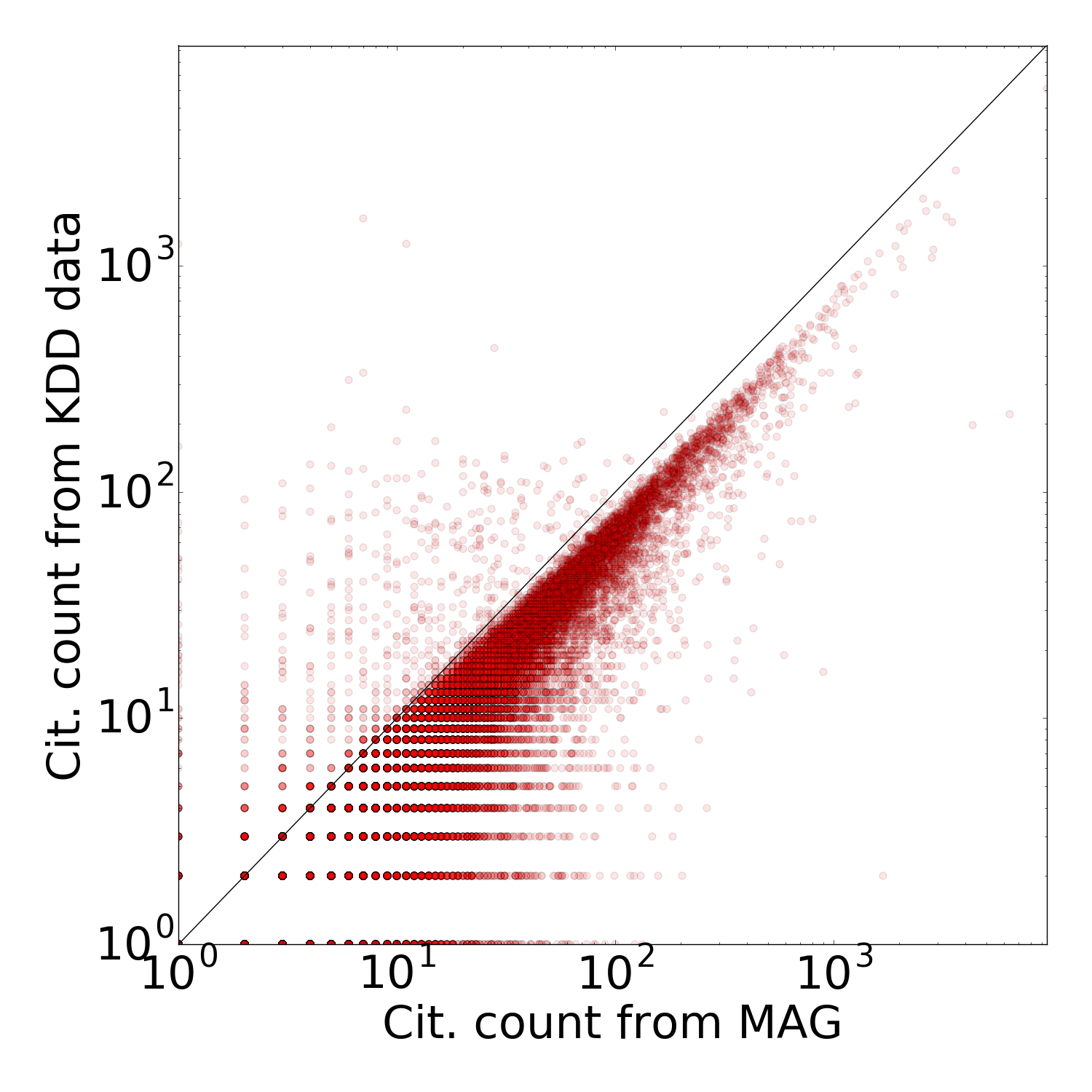}
\caption{Scatter plot of the citation counts reported in the data released for the KDD Cup 2016 and from the online version of the MAG (02/2017).
}
\label{fig:cit_corr}
\end{figure}

For the matched papers ($97.7\%$ of the sample), we compared the number of citations reported in the KDD dump of the MAG with the number of citations reported on the online version of the MAG. 
Since the online MAG has a longer time span than the KDD data, in the absence of noise, we would expect the number of citations in the KDD data to be smaller than or equal to the number of citations reported in the online MAG.
We find that the two citation counts are highly correlated (Fig. \ref{fig:cit_corr}), and only the $2.2\%$ of the sampled papers have more citations in the KDD data compared to the online version of the MAG.
This sets a lower boundary for the error in the percentage of papers with wrong number of citations to about $2.2\%$.

To summarize, after our filtering procedure, we find that the data released for KDD Cup 2016 has about $2.3\%$ of papers with duplicates. 
In addition, about $2.2\%$ of the matched papers have errors in their citation count. 
This means that we have correct citation information for about $95.6\%$ of the analyzed papers. 

\onecolumngrid

\section{Evaluating individual contributions to the Mahalanobis Distance}
\label{app:MD}
The Mahalanobis distance ($d_{\mathcal{M}}$) is an established measure in statistics which generalizes the concept of $z$-score to multivariate distributions by taking into account also possible correlations between the random variables \citep{Mahalanobis1936distance}. 
Its definition reads
\begin{equation}\label{eq:MD}
d_{\mathcal{M}}(\vec{x},\vec{y})=\sqrt{(\vec{x}-\vec{y})^T \mathbf{S}^{-1}(\vec{x}-\vec{y})}
\end{equation}
where $\mathbf{S}^{-1}$ is the inverse of the covariance matrix, $\vec{x}$ and $\vec{y}$ are two vectors containing the random variables. When the covariance matrix is diagonal, i.e. the random variables are not correlated, than the $d_{\mathcal{M}}$ is equivalent to the square root of the sum of the squares of the $z$-scores.

In Sec. \ref{sec:field_bias2}, we have used Eq.~\eqref{eq:contr_md}, an expression for the $d_{\mathcal{M}}$ valid when the covariance matrix comes from a Multivariate Hypergeometric Distribution (MHD), i.e., when the elements of the matrix are
\begin{equation}\label{eq:cov_matrix}
S_{ij}=\left(\delta_{ij}(K_i(N-K_i)) -\left(1-\delta_{ij}\right)K_iK_j \right)\gamma \quad \forall i,j=1,\ldots,F-1
\end{equation}
where $\delta_{ij}$ is the Kronecker delta, $K_i$ is the number of papers of category $i$, $N=\sum_i^F K_i$ is the total number of papers, $F$ is the number of paper categories, $\gamma\!=\!\frac{n(N-n)}{N^2(N-1)}$ and $n$ is the number of sampled papers.
It is worthy to remember that even though we have $F$ different categories, we only have $F-1$ degrees of freedom.
Here, we derive Eq.~\eqref{eq:contr_md} for the case of a MHD in three dimensions, i.e., for $F=3$.
In this case, the covariance matrix is $2\times 2$:
\begin{equation}\label{eg_eq:cov}
\mathbf{S}=\gamma
\begin{pmatrix}
   K_1(N-K_1)      & -K_1K_2 \\
   -K_1K_2       & K_2(N-K_2) \\
\end{pmatrix} \,
\end{equation}
and the inverse of the covariance matrix is
\begin{equation}\label{eq:eg_inv_cov}
\mathbf{S}^{-1}=\frac{1}{\gamma det(\mathbf{S})}
\begin{pmatrix}
   K_2(N-K_2)      & K_1K_2 \\
   K_1K_2       & K_1(N-K_1) \\
\end{pmatrix}
\end{equation}
where $det(\mathbf{S})=K_1(N-K_1)K_2(N-K_2)-(K_1K_2)^2$ denotes the determinant of the covariance matrix, $S$.
Then, let us consider two random column vectors extracted from a $3$-dimensional MHD, $\vec{x}=(x_1,x_2,x_3)^T$ and $\vec{y}=(y_1,y_2,y_3)^T$ such that $n=\sum_{i=1}^3 x_i = \sum_{i=1}^3 y_i$ where $n$ is the number of sampled papers.
Substituting Eq.~\eqref{eq:eg_inv_cov} in Eq.~\eqref{eq:MD}, we write the square of the $d_{\mathcal{M}}$ between $\vec{x}$ and $\vec{y}$ as
\begin{align*}
d_{\mathcal{M}}(\vec{x},\vec{y})^2 = &
\frac{1}{\gamma \, det(\mathbf{S})}
\begin{matrix}
\begin{pmatrix}
x_1-y_1 & x_2-y_2
\end{pmatrix}\\
\mbox{}
\end{matrix}
\begin{pmatrix}  
  K_2(N-K_2) &+K_1K_2 \\
  +K_1K_2  & K_1(N-K_1) 
\end{pmatrix} 
\begin{pmatrix} 
x_1-y_1 \\ 
x_2-y_2 
\end{pmatrix}\\
=&\frac{1}{\gamma \, det(\mathbf{S})}\left\{(x_1 - y_1)^2 K_2(N - K_2)+ (x_2 - y_2)^2 K_1(N - K_1) \right. \\
\,& +\left. 2(x_1 - y_1)(x_2-y_2)(K_1 K_2))\right\}
\\
\,=& \frac{1}{\gamma \, det(\mathbf{S})}\left\{(x_1 - y_1)^2 K_2(K_1 + K_3) + (x_2 - y_2)^2 K_1(K_2 + K_3)  \right.
\\
\,&+\left. 2(x_1 - y_1)(x_2-y_2)(K_1 K_2)\right\}
\\
\,=& \frac{1}{\gamma \, det(\mathbf{S})} \left\{(x_1 - y_1)^2 K_2 K_3 + (x_2 - y_2)^2 K_1 K_3  \right. 
\\
\,&+\left. \left[(x_1 - y_1) + (x_2 - y_2)\right]^2 K_1 K_2 \right\}
\end{align*}
where we have used $N = \sum_{i=1}^3 K_i$. 
Recalling that $n=\sum_{i=1}^3 x_i = \sum_{i=1}^3 y_i$, we know that $(x_1 - y_1) + (x_2-y_2) = (x_3 - y_3)$, so we write
\begin{equation}
d_{\mathcal{M}}(\vec{x},\vec{y})^2 =
\frac{1}{\gamma \, det(\mathbf{S})} \left\{(x_1 - y_1)^2 K_2 K_3 + (x_2 - y_2)^2 K_1 K_3 +(x_3 - y_3)^2 K_1 K_2\right\}
\end{equation}
Then, by using the relation $det(\mathbf{S})=N\prod_{i=1}^3K_i$, we have:
\begin{equation}
d_{\mathcal{M}}(\vec{x},\vec{y})^2 =
\frac{1}{\gamma} \sum_{i=1}^3 \frac{(x_i - y_i)^2}{N K_i};
\end{equation}
noticing from Eq.~\eqref{eq:cov_matrix} that $\gamma = S_{i,i} /( K_i (N - K_i))$, we obtain 
\begin{equation}\label{eq:eg_contr_md}
d_{\mathcal{M}}(\vec{x},\vec{y})^2 =
\sum_{i=1}^3 \frac{(x_i - y_i)^2}{ S_{ii}} \frac{K_i (N - K_i)}{ N K_i } = \sum_{i=1}^3 \frac{(x_i - y_i)^2}{S_{ii}} \left(1 -\frac{K_i }{ N } \right)
\end{equation}
Finally, if we choose one of the two vectors to contain the expected values, $\mu_i$, we re-obtain eq.~\eqref{eq:contr_md} since $(x_i - \mu_i)^2/ S_{ii}=z_i^2$. 
To be precise, the covariance matrix is not defined for $i=3$, however the relation  $\gamma = \sigma_{3}^2 /( K_3 (N - K_3))$ holds and therefore also the final result.

Using Mathematica or similar softwares, it is easy to prove analytically that eq.~\eqref{eq:contr_md} holds for small dimensions. 
We have verified it until 6 dimensions. 
Moreover, we have numerically tested this formula by calculating the $d_{\mathcal{M}}$'s between the ranking vectors of the indicators and the vector of expected values, $\vec{\mu}$, with two different alternative methods: (1) by using Eq.~\eqref{eq:MD}, i.e., by inverting the covariance matrix, and (2) by using the eigenvalue decomposition of the covariance matrix\footnote{The matrix $\mathbf{S}$ is symmetric and it has maximal rank because it is the covariance matrix of a multivariate distribution. 
Therefore, we can diagonalize it, $\mathbf{S}=\mathbf{B}^{-1} \mathbf{D} \mathbf{B}$ where the columns of $\mathbf{B}$ form an orthonormal basis; we can also write $\mathbf{S}^{-1} = \mathbf{B}^{-1} \mathbf{D}^{-1} \mathbf{B} $.
With this, we have $d_{\mathcal{M}}^{2}(\vec{x},\vec{y})=\sum_i^{F-1}c_i/\lambda_i$, where \{$\lambda_i$\} are the eigenvalues of $\mathbf{S}$ and \{$c_i$\} are the coordinates of $\vec{x}-\vec{y}$ in the basis which diagonalizes $\mathbf{S}$, i.e. $c_i=\sum_k^{F-1}(x_k-y_k) B^{-1}_{ki}=\sum_k^{F-1} (x_k-y_k)B_{ik}$ where the last equality comes from the orthonormality of $\mathbf{B}$ which implies $\mathbf{B}^{-1}=\mathbf{B}^{T}$.}.
The results of the three methods were all compatible with each other up to 10 decimal digits.
The advantage of using Eq.~\eqref{eq:contr_md} is that we can calculate the $d_{\mathcal{M}}$ between two arbitrary vectors without dealing with any (computationally slow) matrix inversion or diagonalization, and the
number of needed calculations scales linearly with the number of dimensions. 
Importantly, Eq.~\eqref{eq:contr_md} allows also to assess the individual contribution of each dimension (i.e. of each category) to the $d_{\mathcal{M}}^{2}$.
To our best knowledge, we are the first ones to have derived such explicit formula for $d_{\mathcal{M}}$ when the covariance matrix and the random vectors come from a MHD.

\twocolumngrid

\section{First moment rescaling}
\label{appendix:radicchi}
According to \cite{radicchi2008universality}, the distribution of the $c^{f}$ indicator is log-normal:
\begin{equation}\label{eq:RFC_eq}
F(c^f)dc^f=\frac{1}{\sigma c^f \sqrt{2\pi}}e^{-[\log(c^f)-\mu]^2/2\sigma^2}dc^f
\end{equation}
where $\mu=-\sigma^2/2$ and $\sigma$ is fitted from the data.
When eq.~\eqref{eq:RFC_eq} is verified, then also the distributions of citation count, $c_i$, for all the individual fields, $i$, are lognormal:
\begin{equation}\label{eq:F(c-i)}
F(c_i)dc_i=\frac{1}{\sigma c_i \sqrt{2\pi}}e^{-[\log(c_i)-\log(c_0)- \mu]^2/2\sigma^2}dc_i
\end{equation}
where $c_0$ is the mean of $c_i$. 
For lognormal distributions the variance is proportional to the square of the mean and the constant of proportionality is  $(e^{\sigma^2}-1)$.
From Eq.~\eqref{eq:F(c-i)}, we see that the citation counts $c_i$ are distributed lognormally with mean $e^{\mu+\log{(c_0)}+\sigma^2/2}$ and variance $(e^{\sigma^2}-1)e^{2\mu+2log(c_0)+\sigma^2}$. 
Recalling that $\mu=-\sigma^2/2$, we have that the mean is $c_0$, as it is expected, while the variance becomes $(e^{\sigma^2}-1)c_0^2$.
Thus, when eq.~\eqref{eq:RFC_eq} is verified, the variance of the empirical distribution of the citations for each field has to be proportional to the square of the mean citation count.
Moreover, the constant of proportionality has to be $(e^{\sigma^2}-1)$ for every field and year.

The analytic result just presented is in line with the Eq.~(C.1) given in Appendix C of \citep{mariani2016identification}.
There it is shown that a rescaling procedure based on diving the original score by their first moment works if the ratio between standard deviation and mean is constant. 
In the case of the relative citation ratio, we can calculate analytically such constant using the lognormal distribution and obtain the fitting parameter $\sigma^2$.

\begin{figure}[htb]
\centering \includegraphics[width=0.45\textwidth]{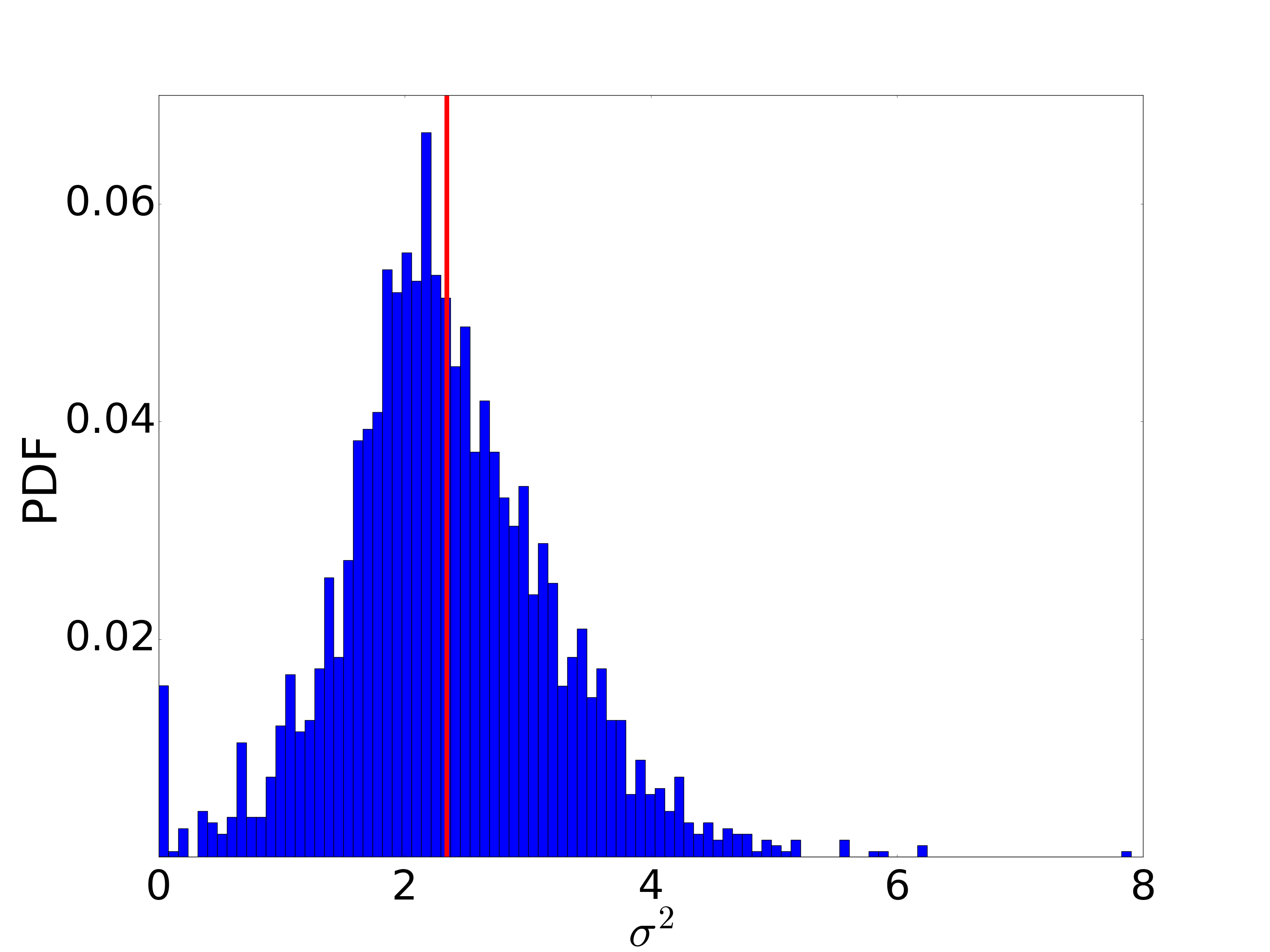}
\caption{Distribution of $\sigma^2$ obtained by calculating the empirical ratio, $r=(e^{\sigma^2}-1)$, between the variance and the square of the mean citation count of each field and year.
}
\label{fig:rfc_universal}
\end{figure}
In Fig.~\ref{fig:rfc_universal}, we report the distribution of $\sigma^2$ obtained by calculating the empirical ratio between the variance and the square of the mean, $r$, and by inverting the relation $r=(e^{\sigma^2}-1)$ for every field and year.
If the universality claim was correct, we would expect a narrow distribution of $\sigma^{2}$. 
By contrast, we find that $\sigma^2$ ranges between $0$ and $8$ across different fields and years.
We argue that the broad range of $\sigma^{2}$ is the reason why the first moment rescaling introduced in \citep{radicchi2008universality} does not work in the analyzed dataset.

\section{Comparing the rankings by the metrics}
\label{sec:comparing}

In Sec.~\ref{sec:quant-rank-bias-by-field-and-age}, we have shown that the rankings by citation count and PageRank can be substantially leveled off across different fields and age groups through a rescaling procedure based on the $z$-score.
The two resulting indicators, $R^{AF}(p)$ and $R^{AF}(c)$, are the least biased among the indicators considered in this paper.
It is important to notice that while citation count has a direct interpretation and it is widely used in research evaluation (\cite{waltman2016review}), PageRank score is a more sophisticated quantity which has not yet be turned into a standard tool for research assessment.
If the rankings by rescaled PageRank and rescaled citation count brought similar information, rescaled citation count might be preferred due to its simpler interpretation and easier computation.

Differently from the citation count, PageRank score uses information on the whole network topology to compute each paper's score. While there exists no universal criterion to decide whether PageRank score leads to a better ranking than the citation count, the results by \cite{chen2007finding} and \cite{mariani2016identification} suggest that in citation networks, PageRank score improves our ability to find groundbreaking publications in the data. 
On the other hand, a node that received many citations is more likely to achieve larger PageRank score -- \cite{fortunato2006approximating} showed that PageRank score is on average proportional to citation count for uncorrelated networks.

\begin{table}[t]
\centering
\begin{tabular}{l*{4}{c}r}
Metrics              & $r$ & $r$ (\text{only} $c>10$) & $\rho$ & $\rho$ (\text{only} $c>10$) \\
\hline
$c,p$ & $0.82$ & $0.82$ & $0.79$ & $0.59$ \\
$R^{AF}(c),R^{AF}(p)$            & $0.80$ & $0.81$ & $0.82$ & $0.74$  \\
\end{tabular}
\caption{Correlations between the metrics. The four columns represent (from left to right): Pearson's correlation coefficient $r$, $r$ restricted to papers that received at least ten citations, Spearman's rank correlation coefficient $\rho$, and $\rho$ restricted to papers that received at least ten citations.}
\label{tab:corr}
\end{table}

We address now the question: \emph{To what extent the rankings by (rescaled) PageRank and (rescaled) citation count differ?}
We focus on two correlations: (1) the correlation between citation count and PageRank, which has been of interest for previous studies (\cite{pandurangan2002using,chen2004local, fortunato2006approximating}) due to the essential role played by PageRank algorithm in determining the success of Google's Web search engine; (2) the correlation between the two rescaled indicators $R^{AF}(c)$ and $R^{AF}(p)$.
The measured correlations are all positive, significantly larger than zero, yet significantly smaller than one (see Table \ref{tab:corr}). 
This is interesting as it may point out that, in analogy with the findings by \cite{chen2007finding} and \cite{mariani2016identification}, network topology brings useful information that is neglected by citation count. 
Whether the additional information used by PageRank metric can be used to identify groups of significant papers will be the subject of future research.

Differently than the rankings by citation count and PageRank score (not shown here), the top papers identified by $R^{AF}(c)$ and $R^{AF}(p)$ come from diverse historical period and diverse fields.
Due to the lack of time bias, also very recent papers can reach the top of the rankings by $R^{AF}(c)$ and $R^{AF}(p)$. 
It would be instructive to look at the top papers as ranked by rescaled citation count and rescaled PageRank. 
However, the original MAG datasets presents noisy entries, as reported by the KDD cup 2016 (see Appendix \ref{app:kdd} for more details), and it causes the scores of some recent papers to be over-estimated, which makes some of the entries in the top-$20$ of the rankings by the rescaled scores unreliable.
For this reason, we do not show the rankings here.
At the same, this problem does not affect the statistical results presented in the previous sections.

\end{document}